\documentclass[reprint,
amsmath,
amssymb,
floatfix,
superscriptaddress,
aps,
prl]{revtex4-2}
\usepackage{graphicx}
\usepackage{dcolumn}
\usepackage{bm}
\usepackage[dvipsnames]{xcolor}
\usepackage{hyperref}%
\hypersetup{
    colorlinks=true,
    linkcolor=MidnightBlue,
    citecolor=MidnightBlue,   
   urlcolor=MidnightBlue,
   }

\def\bea{\begin{eqnarray}}
\def\eea{\end{eqnarray}}
\def\beq{\begin{equation}}
\def\eeq{\end{equation}}

\def\f{\frac}

\def\la{\langle}
\def\ra{\rangle}

\def\nv{ {\hat{n}}}
\def\rv{ {\vec{r}}}
\def\lv{ {\vec{l}}}
\def\fv{ {\vec{f}}}
\def\Fv{ {\vec{F}}}

\def\a{\alpha}
\def\apara{\alpha_{\parallel}}
\def\aperp{\alpha_{\perp}}
\def\b{\beta}

\def\d{\delta}
\def\dtheta{D_{\theta}}
\def\Dpara{D_{\parallel}}
\def\Dperp{D_{\perp}}
\def\zetav{\vec{\zeta}}

\begin{document}
\title{Noise-induced quenched disorder in dense active systems}

\author{Guozheng Lin}%
\email[]{gzhlin@mail.bnu.edu.cn}
\affiliation{School of Systems Science, Beijing Normal University, Beijing, People’s Republic of China}
\author{Zhangang Han}%
\email[]{zhan@mail.bnu.edu.cn}
\affiliation{School of Systems Science, Beijing Normal University, Beijing, People’s Republic of China}
\author{Amir Shee}%
\email[]{amir.shee@northwestern.edu}
\affiliation{Northwestern Institute on Complex Systems and ESAM, Northwestern University, Evanston, IL 60208, United States of America}
\author{Cristián Huepe}%
\email[]{cristian@northwestern.edu}
\affiliation{School of Systems Science, Beijing Normal University, Beijing, People’s Republic of China}
\affiliation{Northwestern Institute on Complex Systems and ESAM, Northwestern University, Evanston, IL 60208, United States of America}
\affiliation{CHuepe Labs, 2713 West Haddon Ave \#1, Chicago, IL 60622, United States of America}
\date{\today}%
\begin{abstract}
We report and characterize the emergence of a noise-induced state of quenched disorder in a generic model describing a dense sheet of active polar disks with non-isotropic rotational and translational dynamics.
In this state, randomly oriented self-propelled disks become jammed, only displaying small fluctuations about their mean positions and headings. 
The quenched disorder phase appears at intermediate noise levels, between the two states that typically define the flocking transition (a standard disordered state that displays continuously changing headings due to rotational diffusion and a polar order state of collective motion).
We find that the angular fluctuations in this dense system follow an Ornstein-Uhlenbeck process leading to retrograde forces that oppose self-propulsion, and determine its properties.
Using this result, we explain the mechanism behind the emergence of the quenched state and compute analytically its critical noise, showing that it matches our numerical simulations.
We argue that this novel type of state could be observed in a broad range of natural and artificial dense active systems with repulsive interactions.
\end{abstract}
\maketitle



Active agents convert stored or ambient energy into mechanical work, injecting it at the smallest scales of the system~\cite{Romanczuk2012, Elgeti2015, Bechinger2016}.
They typically introduce activity through some form of self-propulsion, interact with neighbors via alignment or attraction-repulsion forces, and can be affected by noise. 
Many different models of active systems have been studied in recent years, with multiple parameter combinations, which could have potentially resulted in a variety of regimes and nonequilibrium phases.
Only a few have been identified up to now, however, corresponding to self-organized states with various forms of (polar or nematic) orientational order \cite{Doostmohammadi2018, Guillamat2018, Martin-Gomez2018}, clustering~\cite{Theurkauff2012, Palacci2013}, or phase separation~\cite{Paoluzzi2022, Fily2012}; as well as to disordered states where agents move in randomly changing directions.

One of the most studied phases displaying orientational order is characterized by collective motion, a state in which all agents are aligned and head in a common direction~\cite{Vicsek2012, Speck2016}.
Examples of collective motion can be found in different types of biological systems, including cytoskeleton-motor proteins~\cite{Kron1986, Ndlec1997, Schaller2010}, bacterial colonies~\cite{Keller1971, Dombrowski2004, Zhang2010}, insect swarms~\cite{Buhl2006, Bazazi2012}, bird flocks~\cite{Reynolds1987, Nagy2010}, and fish schools~\cite{Partridge1980, Couzin2002, Herbert-Read2011}. 
It can also develop in artificial systems, such as active colloidal suspensions~\cite{Theurkauff2012}, colloidal rollers~\cite{Bricard2013, Bricard2015}, vibrated polar disks~\cite{Deseigne2010, Deseigne2012}, or robot swarms~\cite{Mori1965, Turgut2008, Ferrante2012, Brambilla2013, Li2019, Petersen2019, Dorigo2020, Oliveri2021}. 
This type of self-organization was originally thought to require local alignment interactions~\cite{Vicsek1995}, but has recently been shown to also emerge from a local coupling between attraction-repulsion forces and heading directions~\cite{Szabo2006}.
Regardless of the underlying mechanism, collective motion corresponds in all these cases to an ordered phase of aligned agents that emerges from a disordered phase with randomly changing headings.
Additionally, both phases are sometimes subdivided into parameter regions with different density distributions~\cite{Huepe2004, Peruani2006, Fily2012, Redner2013, Palacci2013, Cates2013, Cates2015, Reichhardt2015, Zhao2021, ZhaoScientificReports2022}. 
Beyond collective motion, other collective states have been identified more recently in elastic or jammed active solids~\cite{Lin2021, Baconnier2022, Xu2023}.
In these systems, attraction-repulsion forces or steric interactions between densely packed agents can result in different forms of collective oscillations and disordered dynamics \cite{ Henkes2020, Xu2023}.
Despite some initial studies, very little is known about the spatiotemporal states that can develop in active solids or active jamming~\cite{Garcia2015, Henkes2020}.

In this Letter, we report the emergence of a noise-induced state of {\emph{quenched disorder}} (QD) in densely packed active systems, where agents become jammed and their headings fluctuate about different fixed random directions.
%
%
This QD phase appears at intermediate noise levels: For lower noise, most systems self-organize into a state of collective motion that we will refer to as moving order (MO); for higher noise, they reach a standard state of dynamic disorder (DD) where all heading are randomly changing.
We characterize the QD phase in a generic, minimal model of self-propelled disks with off-centered rotation and linear repulsive interactions.
These are similar to the active polar disks with steric interactions introduced in~\cite{Deseigne2010, Deseigne2012}, but with soft repulsive cores and non-isotropic damping.
Using this model, we identify the mechanism that leads to QD, describe it analytically, and show that it could develop in a broad range of systems.

\begin{figure}[!t]
\includegraphics[width=8.6cm]{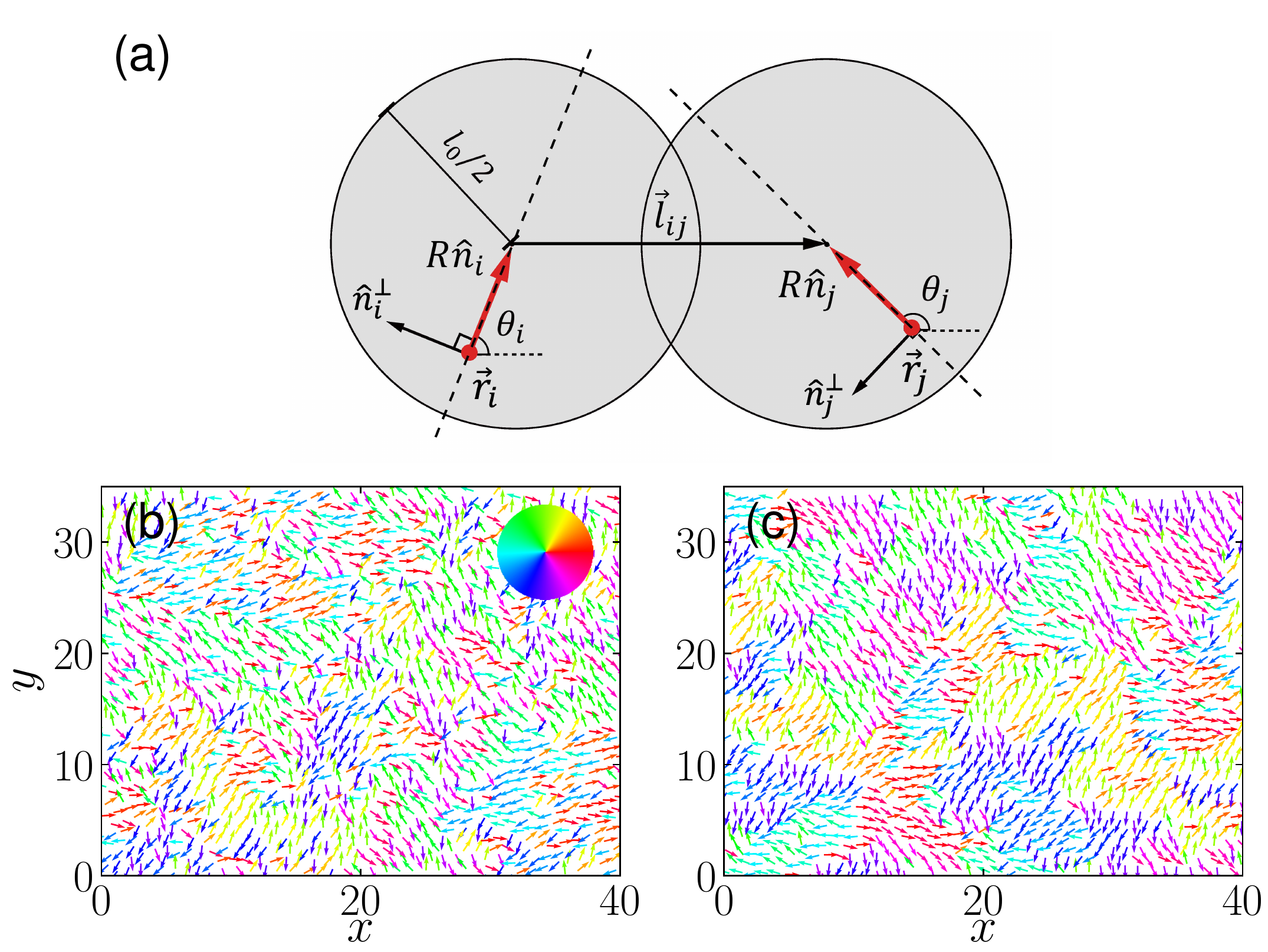}
\caption{(Color online) Schematic representation of model interactions and snapshots of quenched disorder states. Top: Diagram of two soft disks (a), with repulsion radius $l_0/2$ and self-propelled headings, $\nv_i$ and $\nv_j$, translate and rotate about their axes, $\vec{r}_i$ and $\vec{r}_j$, located a distance $R$ behind each centroid.
The linear repulsion, proportional to disk overlap  $\lVert \vec{l}_{ij} \rVert - l_0$, is projected onto each centroid, resulting in forces and torques about $\vec{r}_i$ and $\vec{r}_j$. 
Bottom: Typical quenched disorder state obtained from random initial conditions (b) and a fully aligned initial state (c), both placed on a periodic arena, initially forming a perfect hexagonal lattice. Each disk is represented by an arrow that starts at its centroid, points in its heading direction, and is colored by angle (see inset).
}
\label{fig_1}
\end{figure}     
We consider a system of self-propelled polar disks with rotation axes located behind their geometric centers, interacting through linear repulsive forces.
These can be viewed as minimal representations of self-propelled agents that are nonaxisymmetric about their centers of rotation and thus interact with neighbors anisotropically, which introduces torques.
%
%
Figure \ref{fig_1}(a) illustrates the interactions between two such disks, $i$ and $j$, with radii $l_0/2$ and heading directions $\nv_{i}$ and $\nv_{j}$.
Their axes of rotation $\rv_{i}$ and $\rv_{j}$ are positioned at a distance $0 \leq R \leq l_0/2$ behind their centroids, so $R$ controls the degree of eccentricity of their rotational motion. 
This implies that, for small $R$, the interaction forces will mainly affect disk positions, whereas for large $R$, they will mainly affect their orientations.

We define the interaction between two neighboring disks, $i$ and $j$, as a linear repulsive central force, given by
$\fv_{ij} = k(|\lv_{ij}|-l_0)\lv_{ij}/|\lv_{ij}|$
if $|\lv_{ij}| \leq l_0$,
and by $\fv_{ij}=0$ otherwise.
Here, $k$ determines the repulsion strength and $\rv_{ij}$ is the vector that joins both geometric centers, which can be expressed in terms of the positions of the axes of rotation and the headings as
$\lv_{ij}=(\rv_{j} - \rv_{i}) + R(\nv_{j}-\nv_{i})$.
The total force over disk $i$ is thus given by the sum of pairwise interactions 
$\Fv_i  = \sum_{j}\fv_{ij}$, where $j \in S_i$ is the set of all disks that overlap $i$ (i.e., with center-to-center distance smaller than $l_0$).
Note that, if we add linear attraction forces between neighbors for $|\lv_{ij}| > l_0$, this model would describe the active elastic sheet presented in~\cite{Lin2021}, formed by an hexagonal array of self-propelled rods with front tips permanently linked by linear springs.

By decomposing the effect of the total interactions $\Fv_i$ over the centroid of each disk $i$ into displacement forces and torques about its axis of rotation, we find the following overdamped dynamical equations
\bea
\dot{\rv}_i &=&  v_0 \nv_i + \nv_i \nv^{T}_i\left(\apara\Fv_i + \sqrt{2\Dpara}\zetav_i\right)\nonumber\\  & &+\left(\mathbb{I}-\nv_i \nv^{T}_i\right) \left(\aperp\Fv_i+\sqrt{2\Dperp}\zetav_i\right) ~,
\label{eom1}
\eea
\bea
\dot{\nv}_i &=& \b \left(\mathbb{I}-\nv_i \nv^{T}_i\right)\Fv_i +\sqrt{2 D_\theta} \eta_i (t) \nv^{\perp}_i~.
\label{eom2}
\eea
Here, $v_0$ is the self-propulsion speed and $\nv^{\perp}_i$ is a unit vector perpendicular to $\nv_i$.
Note that, in order to consider a more general model, we included above the possibility of having different damping and noise levels for the disk rotation, front-back translation (along $\nv_i$), and sideways translation (along $\nv^{\perp}_i$).
Rotational motion is thus controlled in Eq.~(\ref{eom2}) by the inverse rotational damping coefficient $\b$ and angular diffusion constant $D_\theta$, whereas translation is controlled in Eq.~(\ref{eom1}) by damping coefficients $\apara$, $\aperp$ and diffusion constants $\Dpara$, $\Dperp$ (along $\nv_i$, $\nv^{\perp}_i$, respectively)~\cite{Ferrante2012, Lin2021}.
Angular noise is introduced through a delta-correlated Gaussian random variable $\eta_i(t)$, with $\la\eta_i\ra= 0$ and $\la\eta_i(t)\eta_j(t^{\prime})\ra = \d_{ij} \d(t-t^{\prime})$.
Positional noise, through a vectorial delta-correlated random variable $\zetav_i(t)$, composed of two independent Gaussian random variables $\zeta^{x}_i(t)$ and $\zeta^{y}_i(t)$, where $\la\zetav_i\ra= 0$ and $\la\zeta^{k}_i(t)\zeta^{l}_j(t^{\prime})\ra = \d_{ij} \d_{kl}\d(t-t^{\prime})$, with indexes $k$ and $l$ representing $x$ or $y$.

We carried out simulations of $N$ self-propelled polar disks, using Euler's method to integrate Eqs.~(\ref{eom1}) and (\ref{eom2}) synchronously for all disks in a periodic rectangular arena of size 
$l_0 \sqrt{N} \times l_0 \sqrt{3 N}/2$.
For $N$ even, this fits exactly $\sqrt{N} \times \sqrt{N}$ disks in a perfect hexagonal lattice with all neighbors at the edge of their repulsive potentials (i.e., with distance $l_0$ between neighboring geometrical centers).
This spatial configuration was used as initial condition, with all angles either aligned in the $x$ direction or selected at random.
As we explored the phase space, we found three possible steady states: MO, DD, and the aforementioned QD state.
States MO and DD have been well documented in the literature, as they correspond to the standard order-disorder (flocking) transition in collective motion. Instead, state QD had not been previously reported and will be the focus of what remains of this Letter. 

Figures \ref{fig_1}(b) and \ref{fig_1}(c) display examples of QD states obtained in simulations.
%
Panel (b) is a snapshot of the stationary solution reached starting from random initial angles, whereas panel (c) shows the state reached starting with all headings aligned.
We observe that the final spatial distribution depends on the initial conditions, as the latter presents larger domains of locally aligned agents.
In both cases, all disks are jammed when the QD state is reached, presenting essentially fixed mean positions and orientations.
Note, however, that some changes may occur due to particle rearrangements, especially in smaller systems, but these will only develop at extremely large timescales.

Our phase space explorations found that QD appears for various combinations of the parameters in Eqs.~(\ref{eom1}) and (\ref{eom2}), as we show in the Supplemental Material \cite{Supply2023}, but not in the often studied case with fully isotropic damping ($\apara = \aperp$), or in limit cases with no angular noise ($\dtheta = 0$) or no rotational anisotropy ($R=0$).
In order to study the emergence of QD in the simplest possible context, we will thus focus on a different limit case, setting $\aperp = \Dpara = \Dperp = 0$, with $\apara>0$ and $\dtheta>0$.
In addition, we will fix in all simulations below $\apara=0.02$, $\b = 1.2$, $k=5$, $l_0 = 1$, $v_0=0.002$, $N=1600$, and $dt=0.01$, while varying $R$, $D_\theta$.

To analyze our simulation results, we introduce two order parameters that allow us to discriminate between the collective states.
The first one corresponds to the standard polarization 
$\psi = \langle||\sum_{i=1}^{N} \nv_i||\rangle_t/N$
(where $\langle\cdot\rangle_t$ is the average over time after reaching a steady state),
which determines the degree of alignment between agents.
%
We thus have $\psi =1$ if all agents are perfectly aligned and $\psi =0$ if they are randomly oriented.
The second one evaluates the persistence of the orientation of each agent over time, averaged over all agents, and is defined by $\phi = \sum_{i=1}^{N} ||\langle\nv_i\rangle_t||/N$.
If the orientation of each agent fluctuates about a fixed mean value, we have $\phi =1$; if they are randomly rotating over time, $\phi = 0$.

\begin{figure}[!t]
\includegraphics[width=8.6cm]{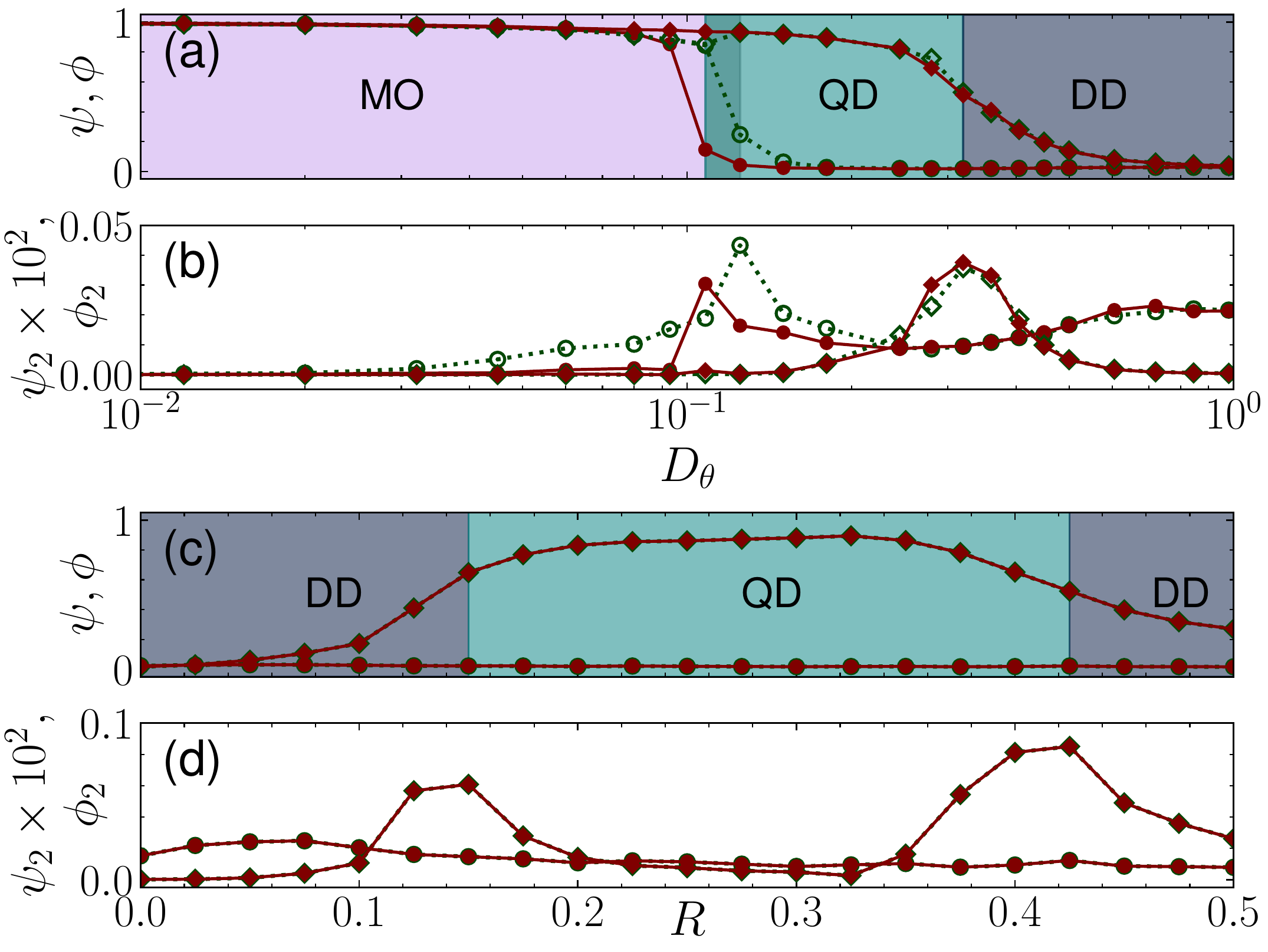}
\caption{(Color online) Order parameters (a,c) and their variances (b,d) as a function of angular noise $\dtheta$ (a,b) and degree of rotational eccentricity $R$ (c,d), for fixed $R=0.3$ or fixed $\dtheta=0.2$, respectively. 
The standard polarization order parameter $\psi$ ($\circ$) and orientation persistence order parameter $\phi$ ($\diamond$) are presented, using solid or open symbols for randomly oriented or aligned initial conditions, respectively.
We identify three regimes: 
a high $\psi$, high $\phi$ moving order (MO) state for $\dtheta\leq 0.108$ in (a);
a low $\psi$, high $\phi$ quenched disorder (QD) state for $0.125 < \dtheta \leq 0.32$ in (a) and $0.15 < R \leq 0.425$ in (c);
and a low $\psi$, low $\phi$ dynamic disorder (DD) state for $\dtheta>0.32$ in (a) and $R \leq 0.15$ or $R > 0.425$ in (c).
Each point is averaged over the last $2\times 10^6$ timesteps (of $2\times 10^7$ total), after reaching the steady state.
All simulations used the parameters detailed in the text.}
\label{fig_2}
\end{figure}

%
Figure \ref{fig_2} presents the three phases obtained in our simulations, as a function either of $\dtheta$ for fixed $R=0.3$ (a,b), or of $R$ for fixed $\dtheta=0.2$ (c,d).
Panels (a) and (c) show the steady state values of $\psi$ ($\circ$) and $\phi$ ($\diamond$), starting from either aligned (open symbols) or random (solid symbols) headings.
Panel (a) shows that we find the MO phase ($\psi \approx \phi \approx 1$) at low $\dtheta$, with agents displaying long-range polar order and a persistent orientation. 
At high $\dtheta$, for either low or high $R$ values, we find the DD phase ($\psi \approx \phi \approx 0$), where headings are continuously randomly changing in any direction.
Finally, at intermediate $\dtheta$ and $R$ values, we find the QD phase ($\psi \approx 0$ and $\phi \approx 1$), where each agent has a (randomly oriented) constant mean heading, about which its instantaneous orientation is fluctuating. 

To help identify the boundaries between the phases, we compute in Figs.~\ref{fig_2}(b,d) the order parameter variances, labeled $\psi_2$ and $\phi_2$.
Their maxima correspond to the transition points used to color the MO, DD, and QD regions in panels (a) and (c).
Panel (a) shows that the QD phase is only found at intermediate noise levels for $R=0.3$, between the MO and DD phases. 
Panel (d) shows that it also requires intermediate $R$ values; if $R$ is too big or too small, the system falls into the DD phase.
Note that the transition between MO and DD occurs at a slightly lower critical $D_\theta$ when simulations are started from a state with random, rather than aligned, headings.

%


We now describe the mechanism that leads to the QD state and postulate an approximate representation of its dynamics that will allow us to describe it analytically.
We begin by noting that, in a densely packed system and for a sufficiently large $R$, the disks will be blocked from rotating by neighboring agents. This implies that the angular fluctuations generated by noise will be constrained by the interparticle repulsive forces.
The tangential component of these forces corresponds to a restitution force that opposes the disks' angular fluctuations while their radial component becomes a retrograde force in the $-\nv_{i}$ direction.
We will show below that the angular dynamics are well described by an Ornstein-Uhlenbeck process~\cite{Karatzas1998} and that the transition from the MO phase to the QD phase will occur when the mean retrograde force matches self-propulsion.

We begin by writing an expression for the effective restitution force that results from the repulsion of neighboring disks.
For small angular fluctuations $\Delta\theta(t)$, about the equilibrium point with $\Delta\theta(t) = 0$, the arc followed by the geometric center of the disk can be approximated by a linear displacement
$\Delta x = R\Delta\theta$. 
In the packed case considered here, the agents will thus approximately feel, in average, a linear restitution force given by $\fv\cdot\nv^{\perp} \approx -(k/c) \Delta x$, where $c$ is a proportionality constant that results from averaging over the multiple configurations of neighbor positions and angular fluctuations.
%
%
If we then replace this expression
into the angular equation of motion (\ref{eom2}), we find that, at first order in $\Delta x \ll 1$, the orientation dynamics reduces to an Ornstein-Uhlenbeck process~\cite{Karatzas1998} described by
\bea
\dot{\Delta\theta} &=& -\f{\beta k R}{c} \Delta\theta + \sqrt{2\dtheta} \eta (t).
\eea
Here, $\eta(t)$ is a random variable that describes a noise with zero mean and variance $\la\eta(t)\eta(t^{\prime})\ra=\d(t-t^{\prime})$. The mean-square fluctuation of the orientation as a function of time thus becomes
\bea
\la\Delta\theta^2 \ra(t) &=& \f{c\dtheta}{\b k R } \left(1- e^{-2\b k R t/c} \right).
\label{eq:orientaion_fluc_evolution}
\eea

\begin{figure}[!t]
\includegraphics[width=8.6cm]{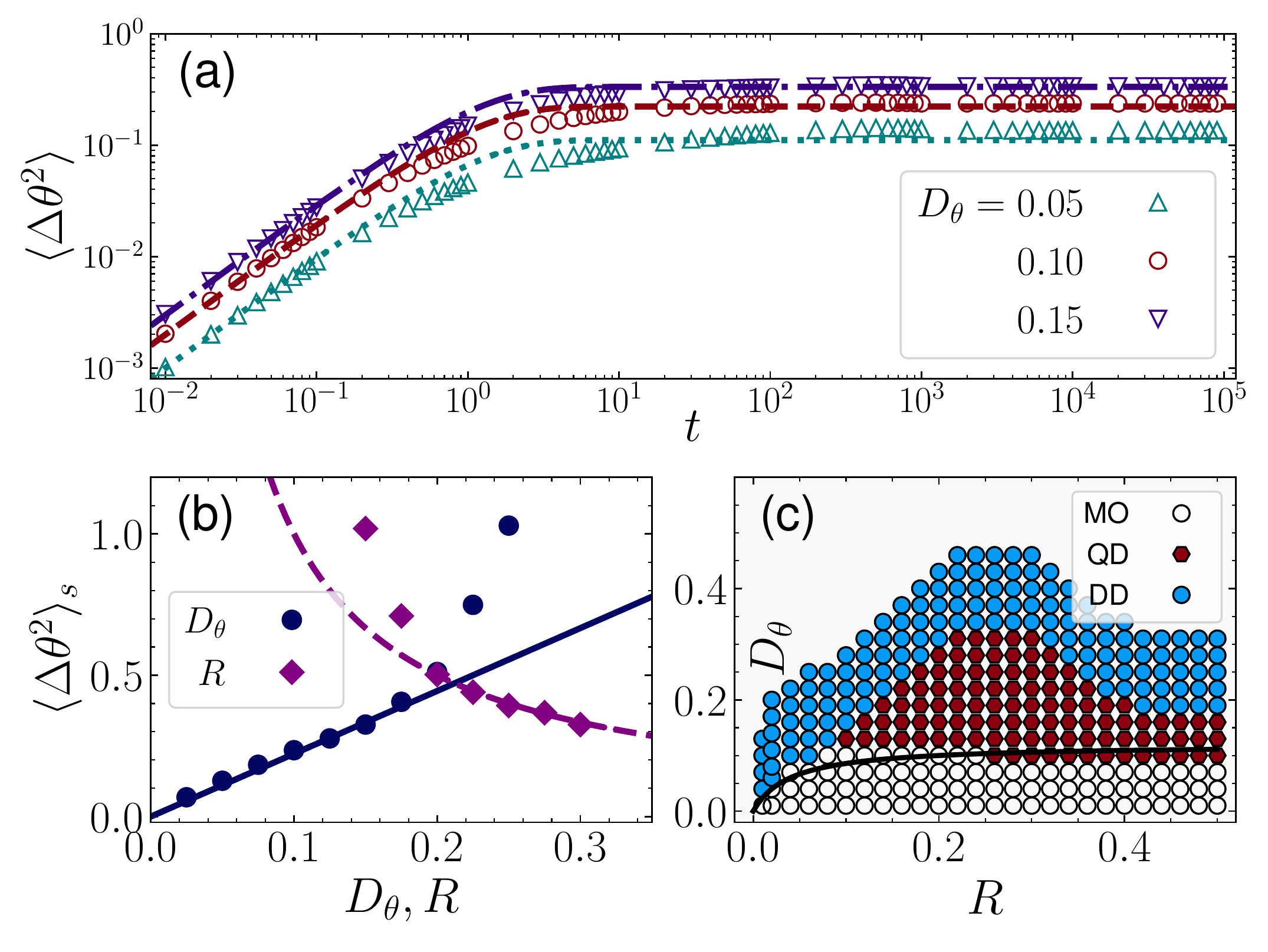}
\caption{(Color online) 
(a) mean-square orientation fluctuations $\la\Delta\theta^2\ra$ as a function of time $t$, for $R=0.3$ and three different values of $D_{\theta}$.
Numerical simulation results (symbols) are well matched by our analytical predictions (curves), expressed in Eq.~(\ref{eq:orientaion_fluc_evolution}), especially in the asymptotic regimes.  
(b) Stead-state mean-square orientation fluctuations as a function of angular noise $D_\theta$ and degree of rotational eccentricity $R$, for fixed $R=0.3$ or fixed $D_\theta=0.15$, respectively.
The numerical simulations and analytical predictions expressed in Eq.~(\ref{eq:steady_state_orientation_fluc}) match well only for low $\la\Delta\theta^2\ra$, as expected.
(c) Phase diagram in the $R$-$\dtheta$ plane. The symbols indicate the phases obtained after reaching the steady state in simulations. No simulations were carried out in the region without symbols, where DD states are expected.
The numerically obtained transition from moving order (MO) to quenched disorder (QD) is well matched by its analytical predictions (solid black curve), expressed in Eq.~(\ref{eq:phase_line}).
All simulations used the parameters detailed in the text and randomly oriented initial conditions.}
\label{fig_3}
\end{figure}

Figure~\ref{fig_3}(a) confirms that, in the MO and the QD state, the mean-square fluctuations of the orientation as a function of time follow our analytical description.
The symbols display the $\la\Delta\theta^2 \ra$ values obtained from numerical simulations; the curves correspond to plots of Eq.~(\ref{eq:orientaion_fluc_evolution}) with $c=4$. 
%
The figure shows that both solutions match very well for the parameters considered in this paper (specified above) and three different $D_{\theta}$ noise levels.
At short timescales, Eq.~(\ref{eq:orientaion_fluc_evolution}) shows that the angular fluctuations in the QD state follow a diffusive behavior with $\la\Delta\theta^2 \ra \simeq 2 D_{\theta} t$.
At long timescales, they saturate at a mean-square value
\bea
\la\Delta\theta^2 \ra_s &=& \f{4\dtheta}{\beta k R}~,
\label{eq:steady_state_orientation_fluc}
\eea
with characteristic crossover time $\tau \sim 4/\b k R$.
%

%

Figure \ref{fig_3}(b) compares the numerical $\la\Delta\theta^2 \ra_s$ values after reaching the steady state,  as a function of $D_\theta$ for fixed $R=0.3$ ($\circ$) and as a function of $R$ for fixed $D_\theta=0.15$ ($\diamond$), to the analytical expression in Eq.~(\ref{eq:steady_state_orientation_fluc}) vs.~$D_\theta$ (solid lines) and vs.~$R$ (dashed lines).
We find an excellent match for $\la\Delta\theta^2 \ra \lesssim 0.5$ but strong deviations for higher $\la\Delta\theta^2\ra$, as expected given our small angle approximations. 

Using the results above, we can determine the transition between the MO and QD phases analytically.
We begin by noting that the repulsive forces that constrain angular fluctuations not only affect $\dot{\Delta\theta}$, but also have a component in the $-\hat{n}$ direction.
%
Defining the mean heading of a disk as $\hat{y} = \la \hat{n} \ra$, we can use Eq.~(\ref{eom1}) to compute the mean force along $\hat{y}$ as
\bea
\la F_{\hat{y}} \ra &=& v_0 \la\nv\cdot\hat{y}\ra+ \a \la(\fv\cdot\nv)(\nv\cdot\hat{y})\ra.
\label{Eq:Fy}
\eea
Since $ \fv\cdot\nv = -k \Delta x \sin(\Delta\theta)/4$ and
$\nv \cdot \hat{y} = \cos \left(\Delta \theta \right)$, we find that
$\la F_{\hat{y}} \ra \approx v_0 - \left( 2 v_0 + \a k R \right) \la \Delta \theta^2 \ra_s / 4$
at leading order in $\Delta x$.
Using Eq.~(\ref{eq:steady_state_orientation_fluc}), we thus obtain
\bea
\la F_{\hat{y}}\ra &\approx& v_0 -\left( 2v_0 + \a k R \right)\f{\dtheta}{\b k R}.
\label{eq:retrogradeF}
\eea
This expression shows that increasing the angular noise $\dtheta$ leads to stronger retrograde forces, which will eventually surpass the self-propulsion term $v_0$ and produce backward motion. When this occurs, the collisions generate anti-alignment forces that result in the quenched state. 
Hence, the critical noise $D_{\theta}^{*}$ can be computed by imposing $\la F_{\hat{y}}\ra = 0$ in Eq.~(\ref{eq:retrogradeF}), which yields
\bea
D_{\theta}^{*} &=& \f{ v_0 \b k R}{2 v_0 + \a k R}.
\label{eq:phase_line}
\eea
%
Figure~\ref{fig_3}(c) shows that the critical noise curve $D_{\theta}^{*}(R)$ matches very well the boundary between the different phases obtained numerically in the $(R,\dtheta)$ plane, for $v_0=0.002$ and all other parameter values specified above, thus validating our assumptions.
The simulation results also show that QD emerges between the MO and DD phases for all $R \gtrsim 0.08$, and that there is an optimal $R \approx 2.26$ at which the QD state remains stable for the highest noise values.
For $R \lesssim 0.08$, angular fluctuations are barely confined since the rotational dynamics are almost isotropic, so the QD mechanisms cannot develop.

In conclusion, our results demonstrate and explain the emergence of a novel, noise-induced QD phase that could be generically present in a broad range of dense active systems.
Although we focused here in a limit case with only angular noise and no sideways displacements, our numerical explorations have shown that the QD state also appears in simulations with a range of nonisotropic damping properties affecting the displacements and with positional noise in addition to the angular noise. 
We emphasize that this is the generally expected situation in real-world systems, where the polar nature of self-propelled agents can be expected to be reflected in anisotropic damping interaction with the substrate. Indeed, various experimental systems have analyzed agents with anisotropic damping~\cite{Deseigne2010, Ferrante2012, Zheng2020, Liu2021, Qi2023, Xu2023}.
Furthermore, in additional to the general model introduced here, we also considered the presence of the QD phase for other models of active agents with repulsive interactions commonly used in the literature \cite{Szabo2006, Henkes2011, Dauchot2019, Baconnier2022}, which are based on self-alignment towards the displacement direction rather than on mechanical torques. We find that QD also emerges (for a range of levels of anisotropy in the noise and the response to external forces) when the angular relaxation is nonlinear \cite{Dauchot2019, Baconnier2022}, but not when it is linear \cite{Szabo2006, Henkes2011} (see Supplemental Material \cite{Supply2023}).
Consequently, we expect the QD state to emerge in experimental systems and encourage the design of setups that could detect it.

This work was  supported by the National Natural Science Foundation of China, Grant 62176022, and by the John Templeton Foundation, Grant 62213. The work of Cristián Huepe was partially funded by CHuepe Labs Inc. The work of Guozheng Lin was partially funded by China Scholarship Council.

\UseRawInputEncoding
\bibliography{reference}
\end{document}